\documentstyle[aps,prl,epsf,amssymb,amsmath,amsthm]{revtex}
\bibstyle{unsrt}

\tighten
\begin{document}
\draft \preprint{}

\twocolumn[\hsize\textwidth\columnwidth\hsize\csname
@twocolumnfalse\endcsname

\title{Complex Extension of Quantum Mechanics}

\author{Carl M. Bender$^1$, Dorje C. Brody$^2$, and Hugh F.
Jones$^2$}

\address{${}^1$Department of Physics, Washington University, St.
Louis, MO 63130, USA}
\address{${}^2$Blackett Laboratory, Imperial College, London SW7
2BZ, UK}

\date{\today}
\maketitle

\begin{abstract}
It is shown that the standard formulation of quantum mechanics in
terms of Hermitian Hamiltonians is overly restrictive. A
consistent physical theory of quantum mechanics can be built on a
complex Hamiltonian that is not Hermitian but satisfies the less
restrictive and more physical condition of space-time reflection
symmetry (${\cal PT}$ symmetry). Thus, there are infinitely many
new Hamiltonians that one can construct to explain experimental
data. One might expect that a quantum theory based on a
non-Hermitian Hamiltonian would violate unitarity. However, if
${\cal PT}$ symmetry is not spontaneously broken, it is possible
to construct a previously unnoticed physical symmetry ${\cal C}$
of the Hamiltonian. Using ${\cal C}$, an inner product is
constructed whose associated norm is positive definite. This
construction is completely general and works for any ${\cal
PT}$-symmetric Hamiltonian. Observables exhibit ${\cal CPT}$
symmetry, and the dynamics is governed by unitary time evolution.
This work is not in conflict with conventional quantum mechanics
but is rather a complex generalization of it.
\end{abstract}

\vskip2pc]

Four years ago it was noted that with properly defined boundary
conditions the spectrum of the Hamiltonian
\begin{equation}
H=p^2+x^2(ix)^\nu\qquad(\nu\geq0) \label{e1}
\end{equation}
is {\em real and positive} \cite{Bender}. The reality of the
spectrum of $H$ is a consequence of its unbroken ${\cal PT}$
symmetry. We say that ${\cal PT}$ symmetry is not spontaneously
broken if eigenfunctions of $H$ are simultaneously eigenfunctions
of ${\cal PT}$. The linear parity operator ${\cal P}$ performs
spatial reflection and has the effect $p\to-p$ and $x\to-x$,
whereas the antilinear time-reversal operator ${\cal T}$ has the
effect $p\to-p$, $x\to x$, and $i\to-i$. While $H$ in (\ref{e1})
is not symmetric under ${\cal P}$ or ${\cal T}$ separately, it is
invariant under their combined operation. We say that such
Hamiltonians possess space-time reflection symmetry. Other
examples \cite{XXX} of complex Hamiltonians having unbroken ${\cal
PT}$ symmetry and thus real, positive spectra are
$H=p^2+x^4(ix)^\nu$ ($\nu>0$), $H=p^2+x^6(ix)^\nu$ ($\nu>0$), and
so on.

Recently, Dorey {\em et al}. proved rigorously that the spectrum
of $H$ in (\ref{e1}) is indeed real and positive \cite{Dorey}.
Many other ${\cal PT}$-symmetric Hamiltonians for which space-time
reflection symmetry is not spontaneously broken have been
investigated, and the spectra of these Hamiltonians have also been
shown to be real and positive \cite{Bender2}.

Nevertheless, if we attempt to develop a consistent quantum theory
for these Hamiltonians, we encounter the severe difficulty of
dealing with Hilbert spaces endowed with indefinite metrics
\cite{Mielnik}. The purpose of the present Letter is to identify a
new symmetry, denoted ${\cal C}$, inherent in all ${\cal
PT}$-symmetric Hamiltonians that possess an unbroken ${\cal
PT}$-symmetry. This allows us to introduce an inner product
structure associated with ${\cal CPT}$ conjugation for which the
norms of quantum states are positive definite. In particular,
${\cal CPT}$ symmetry is shown to generalize the conventional
Hermiticity requirement by replacing it with a dynamically
determined inner product (one that is defined by the Hamiltonian
itself). As a result, we can extend the Hamiltonian and its
eigenstates into the complex domain so that the associated
eigenvalues are real and the underlying dynamics is unitary.

Let us begin by summarizing the mathematical properties of the
solution to the Sturm-Liouville differential equation eigenvalue
problem
\begin{equation}
-\phi_n''(x) + x^2(ix)^\nu \phi_n(x) = E_n \phi_n(x) \label{e2}
\end{equation}
associated with the Hamiltonian $H$ in (\ref{e1}). First, we
emphasize that the differential equation (\ref{e2}) must be
imposed on an infinite contour $\rm{C}$ in the complex-$x$ plane.
For large $|x|$ the contour $\rm{C}$ lies in wedges that are
placed symmetrically with respect to the imaginary-$x$ axis. These
wedges are described in Ref.~\cite{Bender}. The boundary
conditions on the eigenfunctions are that $\phi(x)\to0$
exponentially rapidly as $|x|\to\infty$ on $\rm{C}$. For
$0\leq\nu<2$, the contour $\rm{C}$ may be taken to be the real
axis.

Second, the eigenfunctions $\phi_n(x)$ are simultaneously
eigenstates of the ${\cal PT}$ operator: ${\cal
PT}\phi_n(x)=\lambda_n\phi_n(x)$. Because $({\cal PT})^2=1$ and
${\cal PT}$ involves complex conjugation, it follows that
$|\lambda_n|=1$. Thus, $\lambda_n=e^{i\omega_n}$ is a pure phase.
Without loss of generality for each $n$ this phase can be absorbed
into $\phi_n$ by the multiplicative rescaling $\phi_n \to
e^{-i\omega_n/2}\phi_n$, so that the new eigenvalue is unity:
\begin{equation}
{\cal PT}\phi_n(x)=\phi_n^*(-x)=\phi_n(x). \label{e3}
\end{equation}

Third, the eigenfunctions are complete. The statement of
completeness (for real $x$ and $y$) is
\begin{equation}
\sum_n(-1)^n\phi_n(x)\phi_n(y)=\delta(x-y). \label{e4}
\end{equation}
This is a nontrivial result that has been verified numerically to
extremely high accuracy \cite{Bender4}.

Fourth, there appears to be a natural choice for an inner product
of two functions $f(x)$ and $g(x)$ given by
\begin{equation}
(f,g)\equiv\int_{\rm{C}} dx\,[{\cal PT}f(x)]g(x), \label{e5}
\end{equation}
where ${\cal PT}f(x)= [f(-x)]^*$. The advantage of this inner
product is that the associated norm $(f,f)$ is independent of the
overall phase of $f(x)$ and is conserved in time. Phase
independence is required because we wish to construct a space of
rays to represent quantum mechanical states. With respect to this
inner product the eigenfunctions of $H$ in (\ref{e1}) for all
$\nu\geq0$ satisfy
\begin{equation}
(\phi_m,\phi_n)=(-1)^n \delta_{mn}. \label{e6}
\end{equation}
With this inner product it is easy to verify the position-space
representation of the unity operator in (\ref{e4}). For example,
one can verify that $\int dy\, \delta(x-y)\delta(y-z)=\delta(x-z)$
using (\ref{e4}) and (\ref{e6}). However, because the norms of the
eigenfunctions alternate in sign, the Hilbert space metric
associated with the ${\cal PT}$ inner product $(\cdot,\cdot)$ is
indefinite.

Having reviewed these general properties, the crucial question
that must be addressed is whether a ${\cal PT}$-symmetric
Hamiltonian defines a physically viable quantum mechanics or
whether it merely provides an intriguing Sturm-Liouville problem.
The apparent difficulty with formulating a quantum theory is that
the Hilbert space is spanned by energy eigenstates, of which half
have norm $+1$ and half have norm $-1$. Because the norm of the
states carries a probabilistic interpretation in standard quantum
theory, the existence of an indefinite metric immediately raises
an obstacle.

With the above inner product $(\cdot,\cdot)$, the state space for
a finite-dimensional $2n\times2n$ matrix Hamiltonian that is
symmetric under space-time reflection embodies the symmetry of
$SU(n,n)$ rather than the conventional $SU(2n)$. The space-time
reflection operator in finite dimensions is the product of a
parity operator satisfying ${\cal P}^2=1$ and ${\rm tr}{\cal P}=
0$, and an antilinear Hermitian conjugation operator satisfying
${\cal T}^2=1$. Unlike the unitary group, $SU(n,n)$ is not simply
connected, and consequently the state space is partitioned into
two disjoint halves, one associated with norm $+1$ and the other
with norm $-1$.

One can try to formulate a quantum theory associated with ${\cal
PT}$-symmetric Hamiltonians by insisting that physical states must
have positive norm \cite{Bender3}. This leads to a quantum
mechanics defined on a nonlinear state space. Such investigations
are of interest, but the existence of negative-norm eigenstates
still leaves open serious interpretational issues. In particular,
one finds that expectation values of bounded observables are
unbounded in such quantum theories. The situation here in which
half of the energy eigenstates have positive norm and half have
negative norm is analogous to the problem that Dirac encountered
in formulating the wave equation in relativistic quantum theory
\cite{Dirac}.

We resolve the problem of an indefinite norm by finding a physical
interpretation for the negative norm states. We observe that in
{\em any} theory having an unbroken ${\cal PT}$ symmetry there
exists a previously unnoticed symmetry of the Hamiltonian
connected with the fact that there are equal numbers of
positive-norm and negative-norm states. To describe this symmetry
we construct a linear operator that we denote by ${\cal C}$. We
use the notation ${\cal C}$ because the properties of this
operator are nearly identical to those of the charge conjugation
operator in quantum field theory. The operator ${\cal C}$ is the
observable that represents the measurement of the signature of the
${\cal PT}$ norm of a state. We will see that ${\cal C}$ commutes
with both the Hamiltonian $H$ and the operator ${\cal PT}$.
Therefore, eigenstates of $H$ have definite values of ${\cal C}$.
We will also see that ${\cal C}^2=1$, so the eigenvalues of ${\cal
C}$ are $\pm1$.

We can construct the operator ${\cal C}$ explicitly in terms of
the energy eigenstates of the Hamiltonian. The position-space
representation of ${\cal C}$ is
\begin{equation}
{\cal C}(x,y)=\sum_n\phi_n(x)\phi_n(y). \label{e7}
\end{equation}
Note that ${\cal C}$ is a linear operator. From equations
(\ref{e4}) and (\ref{e6}) we can verify that the square of ${\cal
C}$ is unity:
\begin{equation}
\int dy\,{\cal C}(x,y){\cal C}(y,z)=\delta(x-z). \label{e8}
\end{equation}

We can also construct the parity operator ${\cal P}$ in terms of
the energy eigenstates. In position space
\begin{equation}
{\cal P}(x,y)=\delta(x+y)=\sum_n(-1)^n\phi_n(x)\phi_n(-y).
\label{e9}
\end{equation}
Like the operator ${\cal C}$, the square of the parity operator is
also unity.

The two operators ${\cal P}$ and ${\cal C}$ are distinct square
roots of the unity operator $\delta(x-y)$. That is, while ${\cal
P}^2=1$ and ${\cal C}^2=1$, ${\cal P}$ and ${\cal C}$ are not
identical. Indeed, the parity operator ${\cal P}$ is real, while
${\cal C}$ is complex. Furthermore, these two operators do not
commute; specifically, in the position representation
\begin{equation}
({\cal CP})(x,y)=\sum_n\phi_n(x)\phi_n(-y), \label{e10}
\end{equation}
whereas
\begin{equation}
({\cal PC})(x,y)=\sum_n\phi_n(-x)\phi_n(y), \label{e11}
\end{equation}
which shows by (\ref{e3}) that ${\cal CP}=({\cal PC})^*$.
Evidently, ${\cal C}$ does commute with ${\cal PT}$.

The operator ${\cal C}$ does not exist as a distinct entity in
conventional Hermitian quantum mechanics. Indeed, if we allow the
parameter $\nu$ in (\ref{e1}) to tend to zero, the operator ${\cal
C}$ in this limit becomes identical to ${\cal P}$. Thus, in this
limit the ${\cal CPT}$ operator becomes ${\cal T}$. In short, in
standard quantum mechanics the requirements of ${\cal CPT}$
symmetry and conventional Hermiticity coincide. Therefore, ${\cal
CPT}$ invariance can be viewed as the natural complex extension of
the Hermiticity condition, which ensures the reality of
observables and the unitarity of the dynamics when the Hamiltonian
becomes complex.

Finally, having obtained the operator ${\cal C}$ we can define a
new inner product structure having positive definite signature by
\begin{equation}
\langle f|g\rangle\equiv\int_{\rm{C}} dx\,[{\cal CPT}f(x)]g(x).
\label{e55}
\end{equation}
Like the ${\cal PT}$ inner product (\ref{e5}), this inner product
is also phase independent and conserved in time. The inner product
(\ref{e55}) is positive definite because ${\cal C}$ contributes
$-1$ when it acts on states with negative ${\cal PT}$ norm. In
terms of the ${\cal CPT}$ conjugate, the completeness condition
(\ref{e4}) now reads
\begin{equation}
\sum_n \phi_n(x)[{\cal CPT} \phi_n(y)]=\delta(x-y). \label{e45}
\end{equation}
Unlike the case of conventional quantum mechanics, the ${\cal
CPT}$ inner product is dynamically determined; it implicitly
depends on the choice of Hamiltonian. When $H$ is Hermitian,
(\ref{e45}) reduces to the conventional completeness condition of
quantum mechanics because ${\cal T}$ performs Hermitian
conjugation.

We remark that the ${\cal CPT}$ inner-product (\ref{e55}) defined
above is independent of the choice of integration contour ${\rm
C}$ so long as ${\rm C}$ lies inside the asymptotic wedges
associated with the boundary conditions for the Sturm-Liouville
problem (\ref{e2}). Path independence is a consequence of Cauchy's
theorem and the analyticity of the integrand. In the case of
standard quantum mechanics, where the positive-definite inner
product has the form $\int dx\,f^*(x)g(x)$, the integral must be
taken along the real axis and the path of the integration cannot
be deformed into the complex plane because the integrand is not
analytic. The ${\cal PT}$ inner product (\ref{e5}) shares with
(\ref{e55}) the advantage of analyticity and path independence,
but suffers from the lack of positivity. We find it surprising
that a positive-definite metric structure can be constructed using
${\cal CPT}$ conjugation without disturbing the path independence
of the inner-product integral.

Let us illustrate these results for a $2\times2$ Hamiltonian (see,
also, Ref~\cite{Berry}). In this case we may choose the parity
operator to be
\begin{equation}
\mathcal{P}=\left(\begin{array}{cc} 0 & 1 \cr 1 & 0
\end{array}\right) \label{e12}
\end{equation}
without loss of generality, because in finite dimensions the
parity operator ${\cal P}$ is defined uniquely up to unitary
transformations \cite{Bender3}. Consequently, the most general
form of a ${\cal PT}$-invariant Hamiltonian satisfying ${\cal
P}H^\dagger{\cal P}=H$, where $H^\dagger$ is the Hermitian
conjugate of $H$, can be expressed as a four-parameter family of
matrices
\begin{equation}
H=\left(\begin{array}{cc} re^{i\theta} & s \cr t & re^{-i\theta}
\end{array}\right),
\label{e13}
\end{equation}
where $r$, $s$, $t$, and $\theta$ are real. Note that this
Hamiltonian is not Hermitian in the conventional sense.
Nevertheless, the eigenvalues $\varepsilon_\pm = r \cos\theta \pm
\sqrt{st -r^2 \sin^2\theta} $ are real provided that
\begin{equation}
st>r^2 \sin^2\theta . \label{e15}
\end{equation}
The simultaneous eigenstates of the operators $H$ and ${\cal PT}$
are given by
\begin{equation}
|\varepsilon_+\rangle = \frac{1}{\sqrt{2\cos\alpha}} \left(
\begin{array}{c} e^{i\alpha/2} \cr e^{-i\alpha/2} \end{array}
\right) \label{e16}
\end{equation}
and
\begin{equation}
|\varepsilon_-\rangle = \frac{i}{\sqrt{2\cos\alpha}}
\left(\begin{array}{c} e^{-i\alpha/2} \cr -e^{i\alpha/2}
\end{array}\right), \label{e17}
\end{equation}
where we set $\sin\alpha =(r/\sqrt{st})\,\sin\theta$. It is easy
to verify that $(\varepsilon_{\pm}, \varepsilon_{\pm})=\pm1$ and
that $(\varepsilon_{\pm}, \varepsilon_{\mp})=0$, recalling that
$(u,v)=({\cal PT}u)\cdot v$. Therefore, with respect to the ${\cal
PT}$ inner product, the resulting Hilbert space has a metric
structure of signature $(+,-)$. The condition (\ref{e15}) ensures
that ${\cal PT}$ symmetry is not spontaneously broken. If this
condition is violated, the states (\ref{e16}) and (\ref{e17}) are
no longer eigenstates of ${\cal PT}$ because $\alpha$ becomes
imaginary.

Next, we construct the operator ${\cal C}$. For the parity
operator ${\cal P}$ in (\ref{e12}), the corresponding operator
${\cal C}$ is
\begin{equation}
{\cal C}=\frac{1}{\cos\alpha}\left(\begin{array}{cc} i \sin\alpha
& 1 \cr 1 & -i\sin\alpha \end{array}\right). \label{e18}
\end{equation}
The operator ${\cal C}$ is distinct from $H$ and ${\cal P}$ and
has the key property that
\begin{equation}
{\cal C}|\varepsilon_{\pm}\rangle=\pm |\varepsilon_{\pm}\rangle.
\label{e19}
\end{equation}
The operator ${\cal C}$ commutes with the Hamiltonian and
satisfies ${\cal C}^2 =1$. The eigenvalues of ${\cal C}$ are
precisely the signs of the ${\cal PT}$ norms of the corresponding
eigenstates.

With the aid of the operator ${\cal C}$ we construct the new inner
product structure having positive signature by
\begin{equation}
\langle u|v\rangle=({\cal CPT}u)\cdot v.\label{e20}
\end{equation}
This inner product is positive definite. In particular, we have
$\langle \varepsilon_{\pm}|\varepsilon_{\pm}\rangle=1$. It follows
that the two-dimensional Hilbert space spanned by
$|\varepsilon_\pm\rangle$, with inner product
$\langle\cdot|\cdot\rangle$, has a Hermitian structure with
signature $(+,+)$. Recalling that $\langle u|$ denotes the ${\cal
CPT}$-conjugate of $|u\rangle$, the completeness condition is
\begin{equation}
|\varepsilon_+\rangle\langle\varepsilon_+| +
|\varepsilon_-\rangle\langle \varepsilon_-| =
\left(\begin{array}{cc} 1 & 0 \cr 0 & 1 \end{array} \right).
\label{e21}
\end{equation}
Furthermore, using the ${\cal CPT}$ conjugate
$\langle\varepsilon_\pm|$, we can express ${\cal C}$ in the form $
{\cal C} = |\varepsilon_+\rangle\langle\varepsilon_+| -
|\varepsilon_-\rangle\langle\varepsilon_-|,$ as opposed to the
representation in (\ref{e7}), which uses the ${\cal PT}$
conjugate.

In general, an observable in this theory is represented by a
${\cal CPT}$ invariant operator; that is, one that commutes with
${\cal CPT}$. Thus, if ${\cal CPT}$ symmetry is not spontaneously
broken, the eigenvalues of the observable are real. The operator
${\cal C}$ satisfies this requirement, and hence is an observable.
For the two-state system, if we set $\theta=0$, then the
Hamiltonian in (\ref{e13}) becomes Hermitian in the conventional
sense. However, the operator ${\cal C}$ then reduces to the parity
operator ${\cal P}$. As a consequence, the requirement of ${\cal
CPT}$ invariance reduces to the standard condition of Hermiticity.
It is for this reason that the hidden symmetry ${\cal C}$ was not
noticed previously. The operator ${\cal C}$ only surfaces when we
extend the Hamiltonian of conventional quantum mechanics into the
complex domain.

In summary, we have generalized the condition of Hermiticity in
quantum mechanics to the statement of ${\cal CPT}$ invariance. In
quantum field theory, the conditions of Hermiticity, Lorentz
invariance, and positive spectrum are crucial for establishing
${\cal CPT}$ invariance \cite{Streater}. In this Letter we
establish the converse of the ${\cal CPT}$ theorem in the
following limited sense: We assume that the Hamiltonian possesses
space-time reflection symmetry, and that this symmetry is not
spontaneously broken. From these assumptions, we know that the
spectrum is real and positive and we construct an operator ${\cal
C}$ that is like the charge conjugation operator, and show that
quantum states in this theory have positive norms with respect to
${\cal CPT}$ conjugation. In effect, we replace the mathematical
condition of Hermiticity, whose physical content is somewhat
remote and obscure, by the physical condition of space-time and
charge-conjugation symmetry. These symmetries ensure the reality
of the spectrum of the Hamiltonian in complex quantum theories.

In the real formulation of conventional quantum theory the
dimensionality of the Hilbert space must be even. This is
necessary in order to introduce a complex structure in the
underlying real Hilbert space \cite{Gibbons}. In the present
theory we must introduce the parity operator ${\cal P}$, which can
only be defined on an even-dimensional complex Hilbert space.
Hence, the dimensionality of the underlying real space is a
multiple of four. This is related to the fact that for each
quantum state there is a corresponding partner state, and these
two states are always formed pairwise in the sense that when
${\cal PT}$ symmetry is spontaneously broken, the corresponding
pairs of eigenvalues become complex conjugates of one another.
This is because the secular equation for a ${\cal PT}$-symmetric
Hamiltonian is real \cite{Berry}.

In a conventional Hermitian quantum field theory the operators
${\cal C}$ and ${\cal P}$ satisfy the commutation relation ${\cal
C}{\cal P}=(-1)^N{\cal P}{\cal C}$, where $N$ is the Fermion
number of the state these operators act on. In a ${\cal
PT}$-symmetric quantum field theory the commutation relations are
as follows. If we write ${\cal C}={\cal C}_{\rm R}+i{\cal C}_{\rm
I}$ where ${\cal C}_{\rm R}$ and ${\cal C}_{\rm I}$ are real, then
it follows from (\ref{e10}) and (\ref{e11}) that ${\cal C}_{\rm
R}{\cal P} = {\cal P}{\cal C}_{\rm R}$ and ${\cal C}_{\rm I}{\cal
P}=-{\cal P}{\cal C}_{\rm I}$. As a consequence, it may not
necessarily be true that particles and their partners (for
example, antiparticles) have the same energy eigenvalues. We
recall that space-time reflection symmetry is weaker than the
condition of Hermiticity, and therefore it is possible to consider
new kinds of quantum field theories, whose self-interaction
potentials are, for example, $-ig\varphi^3$ or $-g\varphi^4$, that
have previously been thought to be unacceptable. A plausible
signal of one of these new theories would be the observation of a
particle and its corresponding partner having different masses.

\vskip1pc CMB thanks the U.S.~Department of Energy and DCB thanks
the Royal Society for financial support. The authors are grateful
for helpful remarks from L.~P.~Hughston and T.~W.~B.~Kibble.

\end{document}